# Experimental analysis of DBD plasma jet properties using different gases and two kinds of transfer plate

Fellype do Nascimento[1,a)], Stanislav Moshkalev[1], Munemasa Machida[2]

[1] *Center for Semiconductor Components – State University of Campinas, C. P. 6101, Campinas CEP 13083-870, Brazil*

[2] *Instituto de Física "Gleb Wataghin" – State University of Campinas, Campinas CEP 13083-859, Brazil*

## Abstract

Dielectric Barrier Discharge (DBD) plasma jets has been studied extensively in recent years because of its wide range of applications. DBD plasmas can be produced using many different gases and can be applied to a broad variety of surfaces and substrates. In this work, we provide a comparison of DBD plasmas generated using argon (Ar), helium (He) and nitrogen ($N_2$), as well as their mixtures with water vapor in order to know how some plasma properties are affected by the use of different gases. All plasmas were studied in two different conditions, using a transfer plate made of a conductive material and using a transfer plate made of an insulating one. We observed that the processes of excitation and ionization of nitrogen molecules by direct collisions with Ar or He are more evident and significant in He plasmas, which means that He atoms in metastable states have greater ability to transfer energy to molecules of nitrogen in the plasma. The collisions of He atoms in metastable states with $N_2$ molecules determine the vibrational temperature ($T_{vib}$) values in He plasmas, while in Ar and $N_2$ plasmas the $T_{vib}$ values are determined mainly by collisions of electrons with $N_2$ molecules.

## Introduction

Dielectric Barrier Discharge (DBD) plasma is a non-equilibrium plasma that can be generated at atmospheric pressure, both in open or closed environments [Bibinov2001,Chiper2011]. In this kind of plasma the discharges are produced between two electrodes with at least one of them covered with a dielectric material (glass or ceramic in most cases). DBD plasmas at atmospheric pressure are characterized by low temperature [Masoud2005,Rajasekaran2012,Bashir2014,Machida2015] that is especially important for modification or activation of surfaces of soft materials like polymers or biological tissues, without causing damage to them.

---

a) Corresponding author: fellype@gmail.com

Some authors have been studied the use of argon, helium or nitrogen gases separately in order to produce atmospheric pressure (AP) plasmas [Xu2008,Wei2011,Khatun2010,Muller2013, Hong2008,Walsh2010], some works compared AP plasmas generated using two different gases [Massines2003,Xian2010] and some others studied plasmas formed using gas admixtures [Bibinov2001,Sarani2010]. In this work, we provide a comparison of DBD plasmas generated using three different gases and its admixtures with water vapor: argon (Ar), argon plus water vapor ($Ar+H_2O$), helium (He), helium plus water vapor ($He+H_2O$), nitrogen ($N_2$) and nitrogen plus water vapor ($N_2+H_2O$), in two different conditions: using a transfer plate made of a conductive material (conductive case), and using a transfer plate made of an insulating material (insulating case). Since the transfer plate is equivalent to the sample holders used in many applications of this kind of plasma [Malecha2013,Slepicka2013,Kostov2014,Nascimento2015], it is important to know the influences of the sample holder material in the plasma properties. The comparison of plasmas generated using Ar, He and $N_2$ helped us to understand some mechanisms of energy transfer and the main population processes of excited levels of $N_2$ molecules ($N_2$ I) and $N_2^+$ ions ($N_2$ II) in the Ar, He and $N_2$ plasmas.

To generate the DBD plasma we used a novel device that was built using a 5C22 thyratron valve and a ferrite transformer [Machida2015]. The use of a ferrite transformer allowed an increase of the 5C22 initial pulse duration from ~200 nanoseconds to ~1 microsecond, keeping the initial rise time with a value about 100ns. This results in significantly extended plasma pulses and thus improved efficiency of material processing with the plasma. The device can be operated from 5 to 40kV and can produce from 6 to 300 plasma pulses per second.

## Experimental setup

The setup used to acquire the experimental data used to study the DBD plasma is shown in Figure 1. The part consisting of tubes and electrode represents a transverse section of the DBD reactor.

The device operates as follows: a continuous gas flow is injected inside the PVC tube and high-voltage pulses are applied to the electrode inside the glass tube. A primary discharge is formed in the region between the glass tube and the PVC tube producing a plasma jet that reaches a transfer plate placed at a distance of 5 mm from the end of the PVC tube. The size of the opening in the PVC tube from where the plasma jet is extracted, is 10 mm in diameter. The gas flow rate used to treat PDMS samples was the same for conductive and insulating cases, fixed at 4 liters per minute.

In order to produce plasma discharges using gases with admixture of water vapor, the water vapor is carried by the gas flow using the bubbling method. The vessel with water indicated in the dashed part of Figure 1 is present only in experiments with admixture of water vapor, otherwise the tube leaving the flowmeter is plugged directly to the PVC tube.

Measurements of spectral emissions were carried out using an Andor 303i spectrometer equipped with an iStar DH720 iCCD detector. A 150 lines/mm grating was used in order to get an overview of entire spectra and a 1200 lines/mm grating was used for detailed spectral measurements. The light emitted by the plasmas was collected with a cosine lens and transported to the spectrometer through an optical fiber. The aperture of the entrance slit was set to 250 µm giving us a resolution of (4.45 ± 0.03) nm for the 150 lines/mm grating and (0.418 ± 0.006) nm for the 1200 lines/mm grating.

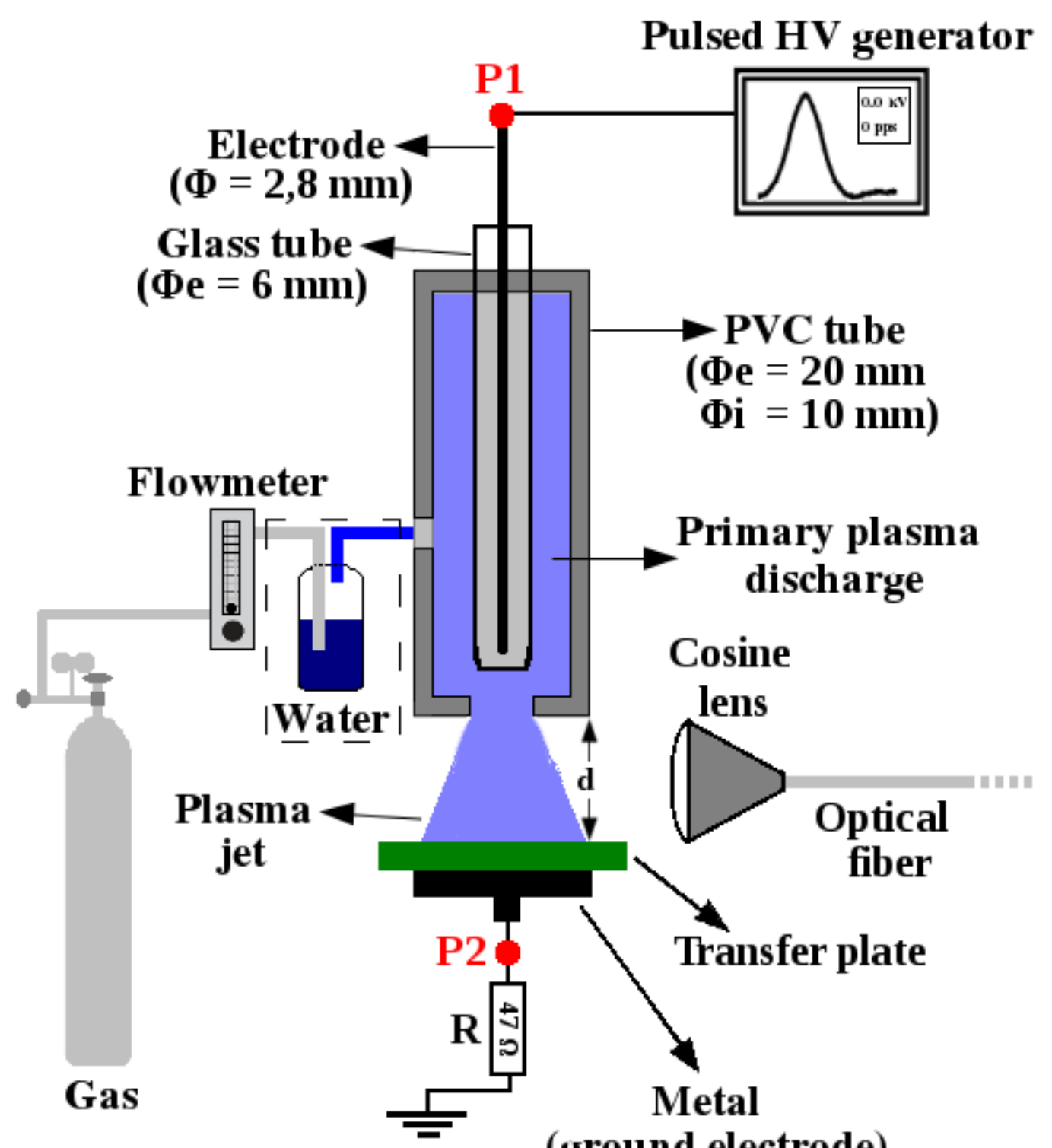

**Figure 1:** Setup used to acquire data to study the DBD plasma. The DBD reactor (PVC tube, glass tube and electrode) is shown in a transverse section. Φe and Φi refer to external and internal diameters, respectively. P1 and P2 are points where voltage probes are connected for voltage and current measurements. The elements are out of scale.

The plasma power estimates were made using a well known method [Machida2015,Ashpis2012] measuring simultaneously the voltage applied on the electrode and the voltage across a shunt resistor $R$ = 47 Ω, which is used to calculate the current using Ohm's law. We can define the value of plasma power as the integration of the product between voltage and current signals during the time of plasma pulse duration multiplied by the pulse repetition rate $f$:

$$P=f\int_{0}^{t} V(t')\,I(t')\,dt' \qquad (1)$$

where $t$ is the duration of the plasma pulse. In other words, we are defining the plasma power as the energy of one plasma pulse multiplied by the number of pulses per second. We used a pulse repetition rate $f$ = 60 Hz for all experiments of this paper.

## Results

The emission spectra of the plasmas produced with Ar, Ar+$H_2O$, He, He+$H_2O$, $N_2$ and $N_2$+$H_2O$ gas flows are shown in Figure 2, for the conductive case, and in Figure 3, for the insulating case. The spectrums in Figures 2 and 3 are accumulations of 1500 plasma pulses acquired using a 150 lines/mm grating.

Emissions from the OH radical (near 308nm) are present in the emission spectra of argon and helium plasmas. Some $N_2$ I molecular line emissions are present in the spectra of all plasmas. The most intense $N_2$ I molecular line emissions appear at 337.1, 357.7 and 380.5 nm. Detectable $N_2$ II molecular line emissions can be seen for helium plasma (at 391.4 and 427.8 nm) and for argon and argon plus water vapor plasmas (at 427.8 nm) in the conductive case.

Atomic line emissions from hydrogen were observed for Ar+$H_2O$ plasma ($H_\alpha$ and $H_\beta$) and for He+$H_2O$ ($H_\alpha$) in the conductive case only. An atomic line emission from oxygen (O I at 777.2 nm) was observed only for Ar+$H_2O$ plasma, in both conductive and insulating cases.

The emission spectra of Ar and Ar+$H_2O$ plasmas show many atomic Ar lines (Ar I at the right end of spectra). When using the He and He+$H_2O$ gases, atomic He line emission (He I at 587.6 nm) appears only for He in the conductive case.

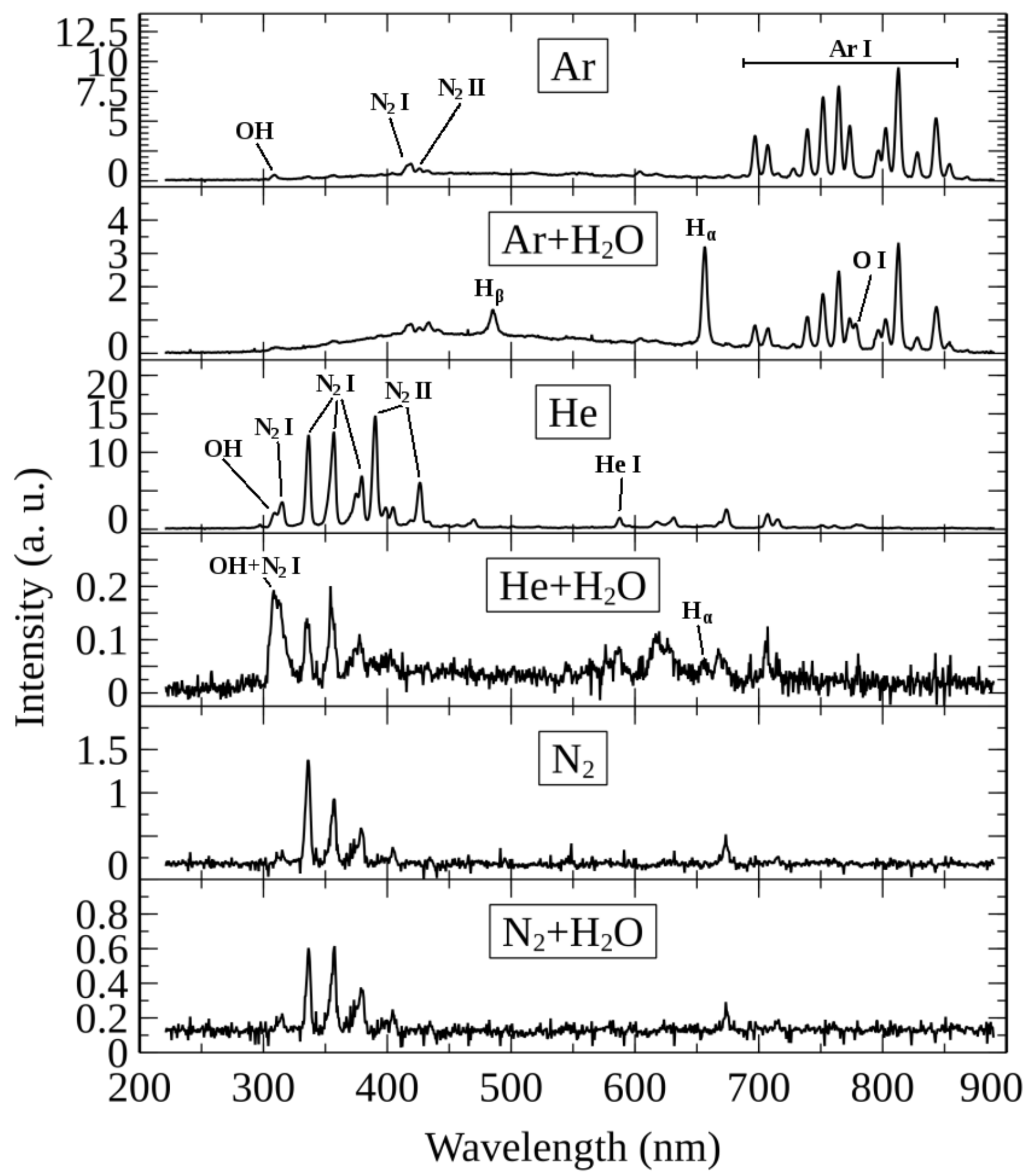

**Figure 2:** Emission spectra obtained for plasmas produced with gases: Ar, Ar+$H_2O$, He, He+$H_2O$, $N_2$ and $N_2$+$H_2O$ using a transfer plate made of a conductive material.

As can be seen in Figures 2 and 3, the use of water vapor results in reduction of intensities of some spectral emissions and complete suppression of others. This may indicate a reduction in the power delivered to the plasma with admixtures of water vapor. The use of water vapor did not increase the intensity of OH light emission, which is in agreement to the observed in another work when there is a high concentration of water vapor in the gas [Sarani2010].

Comparing figures 2 and 3 we can see that the use of a transfer plate made of an insulating material in general results in reduction of intensities of some spectral emissions. But we observed that this reduction of intensities of spectral emissions are not significant for the plasmas formed with $N_2$ and $N_2$+$H_2O$ gases.

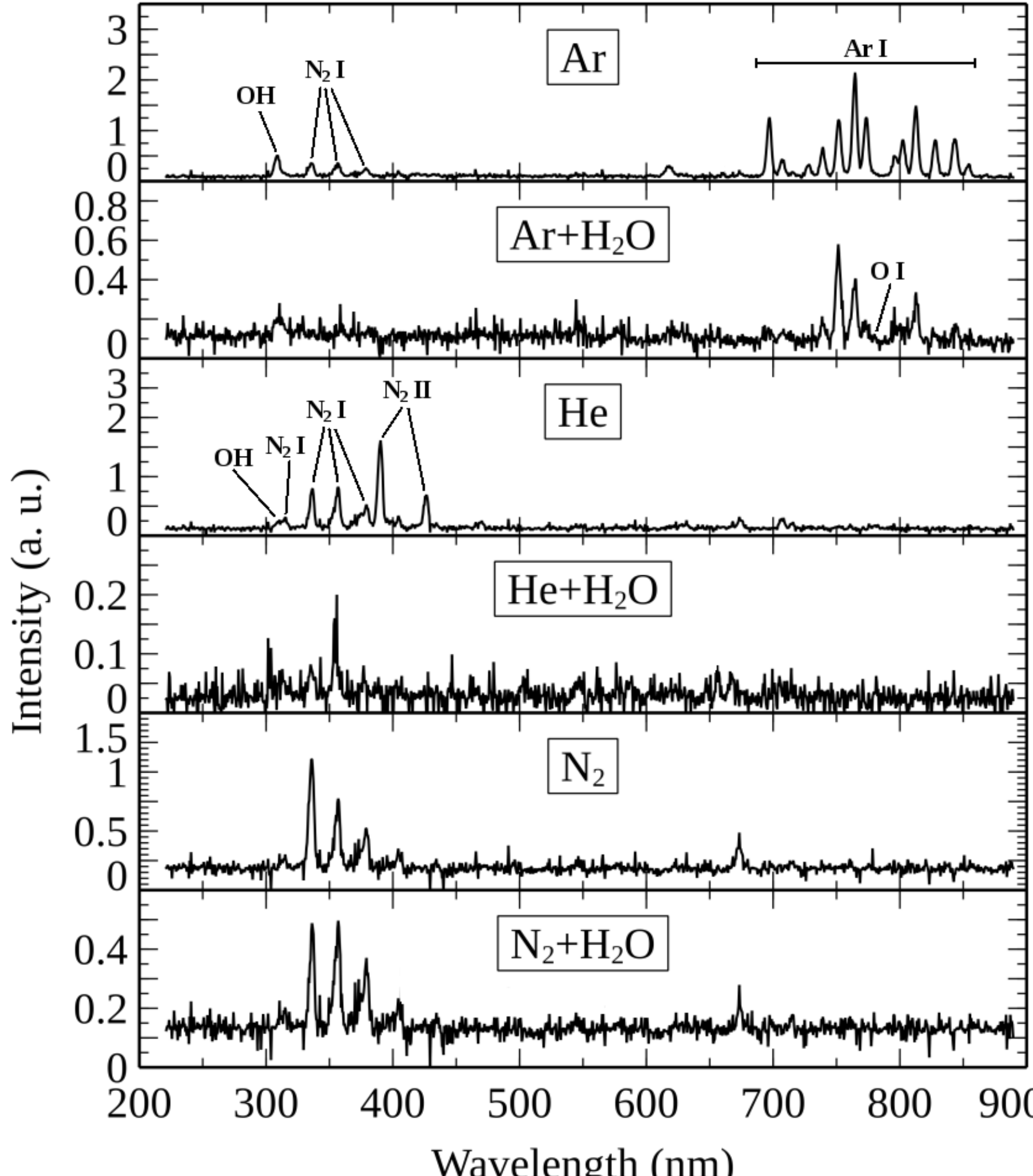

**Figure 3:** Emission spectra obtained for plasmas produced with gases: Ar, Ar+$H_2O$, He, He+$H_2O$, $N_2$ and $N_2$+$H_2O$ using a transfer plate made of an insulating material.

$N_2$ I molecular emission lines (second positive, $C^3\Pi_u$ – $B^3\Pi_g$ transitions) in the spectral range from 365 to 385 nm obtained for plasmas were used to obtain the rotational and vibrational temperatures. The comparison between experimental and simulated curves was used to estimate the rotational temperature, that is known to be approximately equal to the gas temperature ($T_{rot} \approx T_{gas} = T_{plasma}$) [Motret2000,Moon2003,Bruggeman2014], as well as the vibrational temperature ($T_{vib}$) for nitrogen molecules in each case. The simulations were carried out using SpecAir software [refSpecAir]. Table 1 summarizes the results achieved for $T_{rot}$ and $T_{vib}$, with an uncertainty of 10% on each temperature value, in the conductive and insulating cases. As we can see in Table 1, there are no significant changes in the values obtained $T_{rot}$ and $T_{vib}$ when we alternate between the conductive and insulating cases.

**Table 1:** Rotational and vibrational temperatures for $N_2$ for each kind of plasma.

| **Gas** | Insulating transfer plate | | Conductive transfer plate | |
|---|---|---|---|---|
| | **$T_{rot}$ (K)** | **$T_{vib}$ (K)** | **$T_{rot}$ (K)** | **$T_{vib}$ (K)** |
| Ar | 550 | 1800 | 550 | 1800 |
| He | 400 | 2800 | 400 | 2820 |
| $N_2$ | 400 | 2300 | 400 | 2310 |
| $N_2$+$H_2O$ | 400 | 2300 | 400 | 2300 |

It was not possible to determine the $T_{rot}$ and $T_{vib}$ temperatures for Ar+$H_2O$ and He+$H_2O$ plasmas due to too low $N_2$ I emission intensities.

In order to better understand the variation of plasma parameters when different gases are used or when we switch from the conductive to the insulating case, we measured the power delivered to the plasma for all the gases used in this work. Table 2 shows the peak voltage applied to the electrode and the power delivered to each kind of plasma.

**Table 2:** Peak voltage applied to the electrode and power delivered to each kind of plasma.

| **Gas** | **Voltage (kV)** | **Power (mW)** | |
|---|---|---|---|
| | | Insulating | Conductive |
| Ar | 14.0 ± 0.5 | 128.1 ± 9.7 | 664 ± 72 |
| Ar+$H_2O$ | 14.5 ± 0.5 | 40.9 ± 3.1 | 551 ± 45 |
| He | 15.0 ± 0.5 | 77.8 ± 5.8 | 525 ± 39 |
| He+$H_2O$ | 14.5 ± 0.5 | 41.5 ± 2.9 | 133 ± 8.9 |
| $N_2$ | 30.0 ± 0.5 | 106 ± 11 | 195 ± 15 |
| $N_2$+$H_2O$ | 31.5 ± 0.5 | 106 ± 10 | 123 ± 12 |

From the data in Table 2 it can be seen that the use of Ar provides the best relationship between applied voltage and power delivered to the plasma, both in conductive and insulating cases. Ar+$H_2O$ and He gases, in the conductive case, also have a good relationship between applied voltage and power delivered to the plasmas. In general, the addition of water vapor results in reduction of the power delivered to the plasma. The addition of water vapor reduces the electrical conductivity of the plasma, limiting the electric current that passes through it and reducing the power delivered to the plasma.

The voltage needed to ignite $N_2$ and $N_2$+$H_2O$ plasmas are at least twice the voltage needed to ignite the plasmas formed using all other gases. It means that for $N_2$ and $N_2$+$H_2O$ gases, a higher electric field are required to accelerate electrons with enough energy to generate radicals and active species necessary to ionize the gases and ignite the plasmas.

## Discussion

If we take into account the applied voltage used to produce the plasmas and the vibrational temperatures measured for different plasmas, we can obtain information about the energy transfer between electrons and metastable with molecules and relate this to vibrational excitation by electron impact or by metastable impact. Excited species of $N_2$ I can be partly generated by energy transfer from argon and helium metastable, in addition to the electron excitation. The energy carried by the argon and helium metastable are higher than that of energy level of $N_2$ I from the ground state, so the argon and helium metastable can transfer energy to the ground state of $N_2$ I to generate excited $N_2$ I states [Bibinov2001b, Jiang2014]. The $T_{vib}$ values obtained for Ar, $N_2$ and $N_2$+$H_2O$ plasmas looks to be determined by the intensity of the voltage applied on the electrode, which is higher for $N_2$ than for Ar, resulting in a higher $T_{vib}$ value for $N_2$ and $N_2$+$H_2O$ plasmas. The applied voltage influences the energy of free electrons in the plasma, increasing the vibrational excitation by collisions with free electrons. But the $T_{vib}$ values obtained for He plasmas looks to be determined by collisions with energetic metastable. When we compare Ar and He plasma in the conductive case, we can see that the voltage applied to the plasmas are approximately equal and the same occurs for the power delivered to the plasmas, but the $T_{vib}$ values obtained for He plasma is considerably higher than that for Ar plasma.

Looking to the temperatures shown in Table 1 and considering the applied voltage and plasma power shown in Table 2, we can say that the $T_{rot}$ values are not determined by the applied voltage or by the power when different gases are used. For example, the voltage applied to the $N_2$ plasma is twice that applied to the He plasma and, in the conductive case, the power delivered to the He plasma is more than double the power delivered to the $N_2$ plasma, but we obtained the same value fot $T_{rot}$ in both cases. The $T_{rot}$ values are related probably to the thermal conductivity of the gas used.

The main difference between insulating and conductive cases is that the power delivered to the plasma is, in general, much higher in the conductive case. It is simply due to the fact that the electric current passing through the plasma is much higher in that case. Then, it is obvious that we will always have more power in the conductive case. This difference of power delivered to the plasma causes a change in the electron energy distribution function (EEDF), by increasing the number of electrons with more energy. This fact can be noted comparing the spectra obtained with Ar and Ar+$H_2O$ plasmas where we can see that the distribution of intensities of Ar I emission lines changes when the transfer plate is a conductive or an insulating.

The differences in the plasma power also results in the appearance of some spectral emissions like the $N_2$ II for the Ar plasma, $H_\alpha$ and $H_\beta$ for the Ar+$H_2O$ plasma and He I for the He plasma, and changes in the intensities of emissions like Ar I in the Ar and Ar+$H_2O$ plasmas and OH, $N_2$ I and $N_2$ II in the He plasma.

For the He plasma, we can note that the intensities of $N_2$ I lines in relation to $N_2$ II lines are higher in the conductive case. We also note that the distribution of intensities of $N_2$ I lines are practically the same in both cases and the same occurs with $N_2$ II lines. The first fact can be attributed to the changes in the EEDF.

Interesting about the pure nitrogen plasma is that the intensities of $N_2$ I emissions, and also its distribution, do not changed too much when switching from the conductive to the insulating case, as well as the values obtained for $T_{rot}$ and $T_{vib}$. It means that the EEDF and the plasma power do not have strong influence on such parameters for $N_2$ plasma.

The comparison between nitrogen plasma spectrums in the insulating and conductive cases shows us clearly that the intensities of $N_2$ I emissions does not have dependence on the plasma power. The power in the conductive case is almost twice the insulating case, but the intensities of $N_2$ I emissions are practically the same in both cases. The relative intensities of $N_2$ I peaks also remains the same and it indicates that the ohmic heating does not change the EEDF in the $N_2$ plasma.

Another important parameter influencing the plasma behavior and the EEDF is the electric field strength $E$. The reduced electric field strength $E/n$, where $n$ is the gas number density, can be estimated using the ratio between intensities of $N_2$ II and $N_2$ I emissions, I($N_2$ II) and I($N_2$ I) [Paris2005], and a change in this ratio indicates a change in the reduced electric field strength – the higher the ratio I($N_2$ II)/I($N_2$ I), higher the $E/n$ value. Looking to the He plasmas in the Figures 2 and 3, we note that the ratio between intensities of $N_2$ II and $N_2$ I emissions are clearly different in the conductive and insulating cases. For example, in the insulating case the ratio I($N_2$ II, 391 nm)/I($N_2$ I, 337 nm) is twice the same ratio in the conductive case. The intensity of $E$ depends only on the voltage applied to the electrode and the distance between the powered and ground electrodes, and these parameters were kept constants when switching from the conductive to the insulating case. So, we can say that the gas number density is modified when we change from the conductive case to the insulating case, and the gas number density is larger in the conductive case, which reduces the value of $E/n$.

Another interesting point to be noted with the observation of ratios between intensities $N_2$ I and $N_2$ II emissions is when we compare the emission spectra of Ar and He plasmas in the conductive case (Figure 2), the ratio I($N_2$ II, 428 nm)/I($N_2$ I, 419 nm) in the He plasma is much higher than in the Ar plasma. It indicates that the use of different gases also leads to a change in the reduced electric field in the plasma. Note that the applied voltage differs only a few percent for the two gases, therefore the intensity $E$ does not change so much. Then, in this case we also have a reduction in the value of $E/n$ and it means that the gas number density is larger for the Ar plasma.

Figure 4 shows photos of He plasmas in the conductive (a) and insulating (b) cases and also a photo of Ar plasma (c) in the conductive case. In these photos we can see that the plasma volume ($V_P$) are different in all cases, being $V_P(\mathrm{He_{conductive}}) > V_P(\mathrm{He_{insulating}}) > V_P(\mathrm{Ar_{conductive}})$, and this explains the relationship $n(\mathrm{Ar_{conductive}}) > n(\mathrm{He_{insulating}}) > n(\mathrm{He_{conductive}})$.

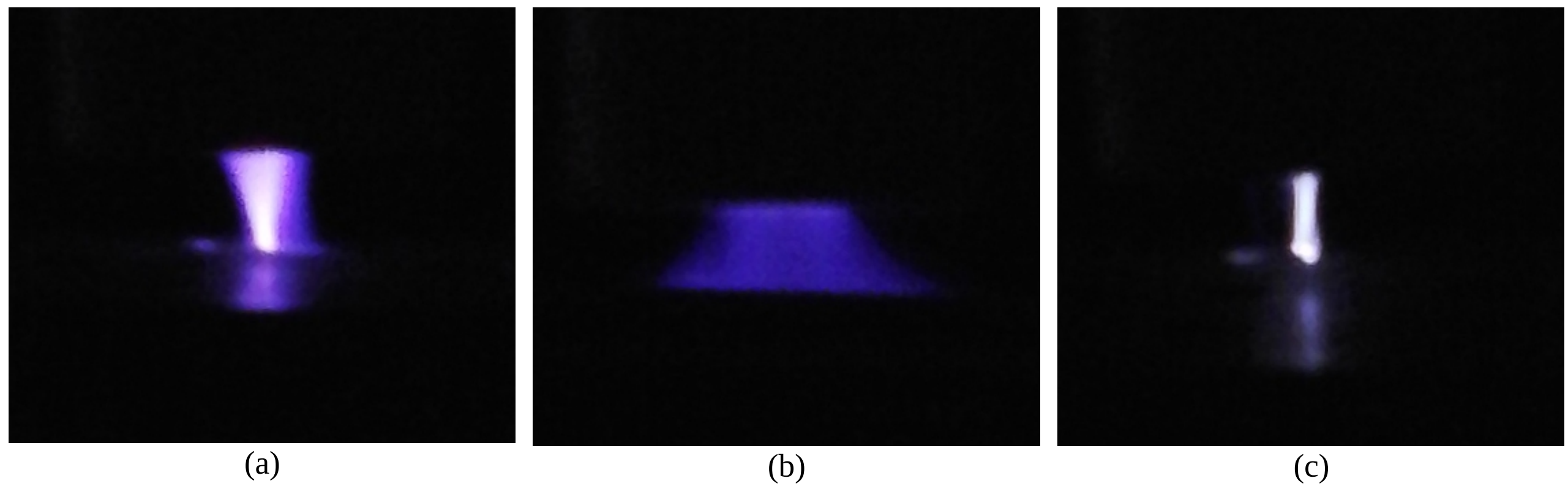

(a) (b) (c)

**Figure 4:** Photos of plasmas: a) He, conductive case; b) He, insulating case; c) Ar, conductive case.

Looking to the distribution of intensities of Ar I lines in Ar and Ar+$H_2O$ plasmas for the conductive case, we can say that the EEDF does not change significantly when water vapor is added to the argon gas. But for all other cases there is a change in the distribution of intensities of all line emission when water vapor is added to a gas. This is due mainly to the reduction of the power delivered to the plasma, followed by a change in the EEDF.

Looking to the intensities of $N_2$ I and $N_2$ II molecular line emissions, we can see that they are higher in the He plasma, in both conductive and insulating cases. This fact indicates that the populations of electrons in the upper levels of $N_2$ I and $N_2$ II are affected by the energy transfer from He metastable.

Penning ionization ($He^M + N_2 \rightarrow N_2^+ + He + e$ or $Ar^M + N_2 \rightarrow N_2^+ + Ar + e$, where M denotes the element is in the metastable state) [Massines2003, Muller2013] is the dominating process in order to produce $N_2^+$ molecular ions for both Ar and He plasmas. There is no evidence of penning ionization in $N_2$ plasma, which indicates that the metastable states of $N_2$ molecules are not able to realize the reaction: $N_2^M + N_2 \rightarrow N_2 + N_2^+ + e$, like occurs in nitrogen afterglow discharges [Guerra2003]. Since in plasmas generated with $N_2$ and $N_2$+$H_2O$ gases there are no other elements in sufficient quantity with energetic metastable colliding with $N_2$ molecules, we can say that the excited levels of $N_2$ I molecular bands are populated only by direct electron impact excitation for these plasmas.

## Conclusions

The changes in the plasma power due to an increase of electric current does not affect the values of $T_{rot}$ and $T_{vib}$. This indicates that ohmic heating does not affects these molecular parameters. This fact indicates that the additional power delivered to the plasma is absorbed by electrons and ions, but not by the molecules.

When an insulating or a conductive material is used as the transfer plate, not only the plasma power and the EEDF are changed, but the reduced electric field strength, and its spatial distribution, is also modified. From the changes in the ratios between intensities of $N_2$ II and $N_2$ I emission lines, we can infer that the use of different gases also leads to modifications on the reduced electric field strength.

Excitation by direct collisions between Ar or He atoms in metastable states with $N_2$ molecules influences the emission spectra of $N_2$ I and $N_2$ II, but this process is more evident and significant in He plasmas. The collisions of He metastable with $N_2$ molecules also determines the vibrational temperature in He plasmas.

It is not the power nor the applied voltage that determines the rotational and vibrational temperatures when different gases are used.

Radicals and active species are generated by collisions with the high energy electrons accelerated by the strong electric field in $N_2$ plasmas.

Further studies are being considered in order to determine the influence of plasma power in translational and electron temperatures of DBD plasmas. The correct measurements of these parameters requires absolute intensity calibration of the spectrometer.

## Acknowledgements

This work was supported by CAPES and CNPq.